# Analytical modelling of stable isotope fractionation of volatile organic compounds in the unsaturated zone

**Daniel Bouchard[1, ‡], Fabien Cornaton[1,#], Patrick Höhener[2], Daniel Hunkeler[1*]**

1) Centre for Hydrogeology and Geothermics, University of Neuchâtel, Rue Emile Argand 11, 2009 Neuchâtel, Switzerland

2) Laboratoire Chimie Provence, UMR 6264, Universités d'Aix-Marseille I, II et III-CNRS, Case 29, 3, Place Victor Hugo, F-13331 Marseille Cedex 3, France

**\***Corresponding author: Centre for Hydrogeology, University of Neuchâtel, Rue Emile-Argand 11, Case postale 158, CH-2009 Neuchâtel, Tel. +41 32 718 2560, Fax + 41 32 718 26 03, daniel.hunkeler@unine.ch

‡ Current address: SNC-Lavalin Environnement Inc., 455 Bld René Lévesque, Montréal, Canada, H2Z 1Z3

# Current address: Water Center for Latin America and the Caribbean, Tecnologico de Monterrey, 64849 Monterrey, N.L. México

**Abstract**

Analytical models were developed that simulate stable isotope ratios of volatile organic compounds (VOCs) near a point source contamination in the unsaturated zone. The models describe diffusive transport of VOCs, biodegradation and source ageing. The mass transport is governed by Fick's law for diffusion, and the equation for reactive transport of VOCs in the soil gas phase was solved for different source geometries and for different boundary conditions. Model results were compared to experimental data from a one-dimensional laboratory column and a radial-symmetric field experiment, and the comparison yielded a satisfying agreement. The model results clearly illustrate the significant isotope fractionation by gas-phase diffusion under transient state conditions. This leads to an initial depletion of heavy isotopes with increasing distance from the source. The isotope evolution of the source is governed by the combined effects of isotope fractionation due to vaporisation, diffusion and biodegradation. The net effect can lead to an enrichment or depletion of the heavy isotope in the remaining organic phase depending on the compound and element considered. Finally, the isotope evolution of molecules migrating away from the source and undergoing degradation is governed by a combined degradation and diffusion isotope effect. This suggests that in the unsaturated zone, the interpretation of biodegradation based on isotope data must always be based on a model combining gas-phase diffusion and degradation.

# 1. Introduction

Compound-specific isotope analysis (CSIA) is increasingly used to assess and sometimes quantify *in situ* biodegradation of organic contaminants in groundwater (Elsner et al., 2005; Hunkeler and Aravena, 2010; Meckenstock et al., 2004; Schmidt et al., 2004 ; Sherwood Lollar et al., 1999). The isotopic method is based on the occurrence of different reaction rates for molecules with light and heavy isotopes, respectively. As a result, molecules with a heavy isotope such as $^{13}C$ ($^{13}C$-molecules) tend to accumulate in the remaining contaminant pool relative to molecules having no heavy isotopes ($^{12}C$-molecules). The shift in isotope ratio and concentration decrease can be linked by the Rayleigh equation (Clark and Fritz, 1997) for saturated zone studies. A growing body of literature has described how isotope data can be used to assess and quantify biodegradation in the saturated zone of the subsurface (Elsner et al., 2005; Meckenstock et al., 2004; Schmidt et al., 2004). The use of stable isotopes for the assessment of natural attenuation of pollutant plumes has become widespread in practice (Hunkeler et al., 2009).

Recently, isotope fractionation of volatile organic compounds (VOCs) has also been investigated under unsaturated conditions in passive vaporisation experiments (Harrington et al., 1999; Kuder et al., 2009), in a column with sandy soil (Bouchard et al., 2008a) and a controlled field experiment (Bouchard et al., 2008b). The field experiment consisted of a non-aqueous phase liquid (NAPL) source of known composition buried in the unsaturated zone which consequently created a radially expanding gas-phase VOC plume (Christophersen et al., 2005). Numerical modelling based on Fick's law was used to understand the mechanisms responsible for the isotope fractionation (Bouchard et al., 2008b). Field measurements of isotope ratios for several compounds and numerical modelling provided evidence for carbon isotope fractionation due to the mass dependence of the diffusion rate and due to differences in biodegradation rates depending on the isotopic composition of molecules (Bouchard et al., 2008b). Molecules including a $^{13}$C atom were found to diffuse slower than molecules with only $^{12}$C atoms, which caused a significant isotope fractionation. The isotope fractionation caused by diffusion was later confirmed by a second experiment, this time conducted in a column packed with alluvial sand (Bouchard et al., 2008a). An artificial NAPL source was emplaced at one end of the column which created a one-dimensional horizontal migration of the gas-phase VOC plume. Additionally, isotope measurements made in the vicinity of the source with time revealed a diffusion-related enrichment of $^{13}$C-molecules at the source. A quantitative relationship between the positive shift at the source and the ratio of the diffusion coefficients of heavy and light compound was found (Bouchard et al., 2008a). The isotope fractionation at the source is caused by different mass fluxes of $^{12}$C-molecules and $^{13}$C-molecules through the porous media. However, in the case of hydrogen isotopes, vapour pressure isotope effects can dominate, leading to an enrichment of light isotopes in the source as shown for methyl *tert*-butyl ether (MTBE) (Kuder et al., 2009). The field and column studies and subsequent modelling succeeded in identifying processes that cause the observed isotope shifts in the unsaturated zone. However, there is a need for analytical models that provide more direct insight into the influence of different processes on isotope ratios. An analytical modelling approach is useful to rapidly simulate the isotope patterns for different isotopes (e.g. $^{13}$C, $^{2}$H, $^{37}$Cl) and source scenarios and hence to explore the possibilities and limits of isotope methods in unsaturated zone studies. Furthermore, analytical models can also aid in evaluating field isotope data as the classical Rayleigh equation does not apply to diffusion controlled systems and as numerical models simulating isotope behaviour in the unsaturated zone are very demanding to apply.

The aims of this study were (i) to develop a mathematical framework for the analytical simulation of the isotope behaviour in the unsaturated zone under transient and steady state conditions, (ii) to evaluate if such a simplified modelling approach can reproduce the main trends of measured isotope data observed in column and field experiments and (iii) to gain additional insight into factors that control isotope fractionation under unsaturated conditions. Two analytical equations were derived, one for a linear and one for a spherical one-dimensional configuration. In addition to simulating the experiments, the analytical solutions were used to evaluate the effect of varying degradation rates and isotope enrichment factors on the evolution of carbon and hydrogen isotope ratios in VOCs. For this part of the study, the one-dimensional model was used which can be considered to represent a common contamination scenario consisting of an NAPL pool floating on the water table from which upward diffusion takes place. Using this approach it also was evaluated whether isotope data can be used to identify and quantify biodegradation under unsaturated conditions.

## 2. Modelling approach

**2.1 Process representation in model.** To facilitate the use of an analytical modelling approach, only a limited number of key processes were taken into account including diffusive transport, equilibrium partitioning and first order degradation. The diffusive transport of dilute VOCs in the unsaturated zone is traditionally modelled by Fick's law (Abriola and Pinder, 1985; Gaganis et al., 2004; Jury et al., 1983; Karapanagioti et al., 2004) and the model described hereafter relies on Fickian diffusion as well. The validity of the law is restricted in some cases, such as for vapours close to trichloroethene or gasoline sources, which can be too dense to assume diluted conditions as requires Fick's law. In that case, the application of the Stefan-Maxwell equations may be necessary for modelling gas-phase diffusion (Baehr and Bruell, 1990; Van de Steene and Verplancke, 2006). Although some studies indicate that density-driven advection can play a role (Falta et al., 1989), this process was neglected. Gas-phase diffusion also largely dominates over aqueous-phase diffusion (Schwarzenbach et al., 2003), and the latter is neglected. The gas diffusion coefficient in a porous medium depends on the mean path length travelled by VOCs, termed the tortuosity factor ($\tau$), the physical properties of the compound and the volume of voids. Because only the mass fraction of organic molecules in the gas phase is relevant for significant mass transport, partitioning of the compound occurring between the different phases composing the system (air, water, solid and liquid NAPL if present) becomes very important. To simplify the model, instantaneous and linear air-water partitioning was assumed. The sandy soils modelled here had low organic carbon content and partitioning from the water to solids had been shown to be negligible (Werner and Höhener, 2003). Also, interfacial partitioning was neglected since it occurs mainly in soils with silt or clay (Hoff et al., 1993). The mathematical notation of partitioning and gas-phase diffusion is given in the Supporting Information (equations S1 + S2). Note that there is a difference between steady-state and transient diffusion. Under a steady-state diffusion regime, concentrations in soil water and on solids are constant (partitioning processes in equilibrium) and VOC diffusion is now only affected by the tortuosity and the volume of voids (Werner and Hohener, 2003) while under transient conditions, diffusion is influenced by partitioning (Grathwohl, 1998; Werner and Hohener, 2003).

Biodegradation of VOCs was also included in the model. In the unsaturated zone, biodegradation of VOC occurs only in the aqueous phase and can be described by a variety of kinetic models (zero-order, first-order, Monod kinetics, or instantaneous bimolecular reaction). The latter two apply to conditions of low oxygen partial pressure and reaction fronts of hydrocarbons with oxygen (Davis et al., 2009; Davis et al., 2005). For fully aerobic conditions, Höhener et al. (2003) gave an outline on how the kinetics of biodegradation is related to the VOC concentration in the gas phase, and how Henry's law is involved therein. Furthermore, it was shown that models based on first-order kinetics were reasonably reproducing the observed vapour profiles in sandy soil columns (Hohener et al., 2003; Jin et al., 1994), in a lysimeter study (Pasteris et al., 2002) and in the field (Hohener et al., 2006).

Several modelling studies that incorporated the processes described above successfully reproduced VOC concentrations observed in a column (Jin et al., 1994), a lysimeter (Karapanagioti et al., 2004) and a field experiment (Gaganis et al., 2004) demonstrating the validity of the approach.

**2.2 Simulated scenarios**

The analytical modelling approach was tested by simulating the isotope evolution observed during a column experiment (Bouchard et al., 2008a) and a field experiment (Bouchard et al., 2008b). A brief description of the experimental set up is provided here and the reader is referred to the published articles for more details. The experiments represent different source geometries of NAPL in the unsaturated zone as described below. The analytical models simulate three chronological states for the VOC evolution observed during the experiments, which are a transient state soon after source emplacement, an equilibrium state and a second transient state due to source depletion. Contaminant vapours are assumed to diffuse through a homogeneous and isotropic porous media and simulations were conducted at temperatures representative for the experiments.

2.2.1. Modelling of isotope evolution during column experiment

In the column experiment, a vessel containing a source of NAPL source (mixture of 10 hydrocarbons, composition given in Bouchard et al. (2008a) was emplaced at one end of a sand packed 1 m long steel cylinder with the other end open to the atmosphere. Sampling ports positioned along the column allowed for the sampling of the VOC vapours diffusing from the source to the atmosphere. This experimental set up is analogue to a large NAPL pool which has accumulated on the water table and from which diffusion to the atmosphere occurs (Figure 1). The VOCs released from the contaminant source are diffusing upwards through the unsaturated zone where biodegradation takes place and may enter the atmosphere in the case of incomplete biodegradation. This typical spill scenario for light NAPL was studied by several researchers investigating the fate of VOCs in the unsaturated zone under field conditions (Franzmann et al., 1999; Hers et al., 2000; Lahvis et al., 1999; Ostendorf and Kampbell, 1991; Smith et al., 1996).

2.2.2. Modelling of isotope evolution during field experiment

The Værløse field experiment consisted of a spherical source of an artificial NAPL (mixture of 14 hydrocarbons, composition given in Broholm et al, 2005) buried in the unsaturated zone from which VOCs were released and underwent biodegradation as they migrated through the unsaturated zone (Christophersen et al., 2005; Hohener et al., 2006). The partial pressure of oxygen was never below 0.18 atm indicating that biodegradation was not oxygen limited. The scenario may represent a NAPL e.g. leaked from an underground storage tank that is held in place by capillary forces (Figure 1). Data from pore gas samplers located at 1 m below the ground surface similarly as the source and at distances between 0 and 5m form the source were used. A similar source scenario was also investigated in a previous field study (Conant et al., 1996).

2.2.3. Evaluating the effect of biodegradation using one-dimensional scenarios

Additional calculations were performed to evaluate the effect of biodegradation on the isotope evolution in more detail using a one-dimensional model. Steady state isotope profiles were simulated for a constant source concentration and variable biodegradation rates (0 $d^{-1}$, 0.1 $d^{-1}$ and 1 $d^{-1}$) and using isotope enrichment factors of -2.3‰ and -1‰, respectively. In addition, the isotope evolution of the source during depletion of a compound from a multicomponent NAPL was evaluated for carbon and hydrogen isotopes using scenarios with and without biodegradation.

**2.3 Isotope modelling**

Model simulations were performed using an analytical reactive transport model. The mathematical formulation is developed in detail in the Supporting Information. The model incorporates isotope fractionation during diffusion and biodegradation. Isotope fractionation during sorption was neglected since laboratory studies for carbon isotopes indicated that it is very small (Schüth et al., 2003). During equilibrium liquid/vapour partitioning, several studies (Harrington et al., 1999; Slater et al., 1999) observed small isotope fractionation for carbon in aromatic compounds, whereas for hydrogen, these effects can be considerable (Wang and Huang, 2003). We thus incorporate isotope fractionation during equilibrium vapour-liquid partitioning in our model.

The model is based on the same principles as the numerical transport model which has previously been used to simulate VOC transport in the field experiment conducted at the Værløse Airbase (Bouchard et al., 2008b). The isotope evolution was simulated by treating the molecules with a different isotopic composition as two separate species, and by attributing to each subspecies a different initial concentration, diffusion coefficient, degradation rate and source depletion rate as explained below. The simulated quantities of the two subspecies were then combined to calculate the isotope ratio. Previous work based on numerical modelling (Bouchard et al., 2008b) has shown that the two-species approach yields results that are close to the three-species approach, which is theoretically needed for compounds that include reactive and non-reactive positions. More details on the two- and three-species approaches are given in the Supporting Information.

The governing partial differential equation for diffusive transport of VOC in soil air with first-order biodegradation in soil water and accounting for the assumptions outlined above was taken from (Werner and Hohener, 2003):

1  $$\frac{\partial C_a}{\partial t} = \nabla \cdot (f_a \tau D_m \nabla C_a) - f_w k_w C_a \qquad (1)$$

2  where $f_w$ is the fraction of VOCs in the water phase and $k_w$ is the biodegradation rate in the

3  water. The complete definition and notation of $f_a$, $f_w$ and other parameters in equation 1 is

4  given in the Supporting Information. All the model parameters including coefficients of the

5  soil physical parameters used in the model are listed in Tables 1, 2 and 3.

6  2.3.1 Initial concentration

7  Concentrations of each compound in the vapour and measurements of initial $\delta^{13}C$ values were

8  used to determine initial concentrations of each subspecies. The initial isotope ratio $^{13}C/^{12}C$

9  ($R_{atom}$) was converted into molecule ratio $^{13}C$-molecule/$^{12}C$molecule ($R_{molecule}$) using:

10  $$R_{molecule} = \frac{R_{atom}}{\left(\frac{1-(n-1(R_{atom}))}{n}\right)} \qquad (2)$$

11  where $n$ is the number of C in the molecule. Then, the initial concentration ($C_0$) is distributed

12  to subspecies ($^l C_0$ and $^h C_0$) according to $R_{molecule}$. For the field study, the initial mass was

13  varied to improve the agreement between measured and simulated concentrations.

14  2.3.2 Diffusion coefficients

15  Due to the greater molecular weight, molecules containing one $^{13}C$ are expected to diffuse

16  slower and hence isotope-specific diffusion coefficients were calculated. The molecular

17  diffusion coefficient obtained from the literature was attributed to the dominant light

18  subspecies ($^l D$). The molecular diffusion coefficient for molecule with $^{13}C$ was derived from

19  $^l D$ (Craig, 1953; Jost, 1960) according to equation 3 (Cerling et al., 1991) and was attributed

20  to the heavy subspecies ($^h D$):

1   $$\frac{^lD}{^hD} = \sqrt{\frac{^hM_w(^lM_w + M_a)}{^lM_w(^hM_w + M_a)}} \qquad (3)$$

2   where $^lM_w$ and $^hM_w$ are the atomic masses of subspecies I (no $^{13}$C) and II (one $^{13}$C),

3   respectively. $M_a$ is the average mass of nitrogen, oxygen and hydrocarbons in the air,

4   calculated using the initial hydrocarbon concentration at the source.

5   2.3.3 Biodegradation rates

6   Analytical modelling made use of independently estimated biodegradation rate constants

7   derived from concentration data in the column (Hohener et al., 2003) and field (Hohener et

8   al., 2006) experiment. The overall biodegradation rate constant was attributed to $^{12}$C-

9   molecules ($^lk$). The smaller biodegradation rate constant for $^{13}$C-molecules ($^hk$) was

10  determined using:

11  $\alpha = {^hk}/{^lk}$ \qquad (4)

12  Where $\alpha$ is the isotope fractionation factor for the compound of interest taken from

13  (Bouchard et al., 2008c) and (Bouchard et al., 2005). For the column study, the

14  biodegradation rate constant from the previous study was used as an initial value that requires

15  adjustment as the biomass composition might have changed.

16  2.3.4 Source depletion

17  As compounds are removed from the NAPL source, the isotope ratio of the compounds shift

18  due to isotope fractionation associated with vaporisation and diffusion through the porous

19  medium. In addition, degradation indirectly affects the isotope evolution by modifying the

20  concentration gradients at the boundary of the source. A mathematical expression was

21  derived that incorporates the effect of vaporisation, diffusion and biodegradation on the

22  source isotope evolution (see Supporting Information). The following expression for the source isotope fractionation factor is obtained that relates the rate of removal of light and heavy subspecies for the one-dimensional system:

$$\alpha_{source} = \frac{^h\beta}{^l\beta} = \Psi\sqrt{\alpha}\sqrt{\frac{^hD_m}{^lD_m}}\frac{^hP}{^lP} \tag{5a}$$

with

$$\Psi = \frac{\sinh\left(\sqrt{\frac{^lk}{^lD}}L\right)\cosh\left(\sqrt{\frac{^hk}{^hD}}L\right)}{\sinh\left(\sqrt{\frac{^hk}{^hD}}L\right)\cosh\left(\sqrt{\frac{^lk}{^lD}}L\right)}$$

For radial symmetric diffusion, the source fractionation factor is given by:

$$\alpha_{source} = \frac{^h\beta}{^l\beta} = \frac{^hD_e}{^lD_e}\cdot\frac{^hP}{^lP}\cdot\frac{\left[1+\sqrt{^hk/^hD_e}\cdot r_0\right]}{\left[1+\sqrt{^lk/^lD_e}\cdot r_0\right]} \tag{5b}$$

where $^h\beta$ and $^l\beta$ are source depletion rate constants of heavy and light subspecies, respectively; $^hD_m$ and $^lD_m$ are the diffusion coefficients for heavy and light subspecies, respectively; $^hC_{sat}$ and $^lC_{sat}$ are the gas phase equilibrium concentrations at the source of heavy and light subspecies, respectively; $^hP$ and $^lP$ the vapour pressures of heavy and light subspecies, respectively; $^hk$ and $^lk$ the first order degradation rate of heavy and light subspecies, respectively; $L$ the length of the one-dimensional system and $r_0$ the radius of the spherical source. For the one-dimensional case, $\Psi$ approaches 1 when $L$ is large relative to the square root of k/D. Using typical values of k and D in eq. 5a, one finds that $\Psi$ is equal to 1 for L > 3 meters. A closer look at equation 5 reveals that for compounds where the vapour pressure isotope effect $^hP/^lP$ is small (e.g. for $^{13}C/^{12}C$ in hydrocarbons), the isotope effect associated with biodegradation and diffusion govern the source isotope fractionation factor

1  $\alpha_{source}$. However, for hydrogen isotopes the vapour pressure isotope effect dominates over the
2  two others.
3  Equation 5 makes it possible to calculate the source decay rate constants for the heavy
4  subspecies if the rate of the light subspecies is known. For the column experiment, the source
5  decay rate constants were quantified based on the concentrations measured at the source. For
6  *n*-hexane and benzene the observed behaviour was approximated by a period with constant
7  source until $t = 96h$, followed by source decay. For the field experiment, source decay rate
8  constants from (Broholm et al., 2005) were used that are based on the measured and modelled
9  change in NAPL composition.

10  **3 Results**
11  Measured and modelled $\delta^{13}C$ values of selected compounds in the column (Bouchard et al.,
12  2008a) and field experiment (Bouchard et al., 2008b) are shown in Figures 2 and 3,
13  respectively. The relationship between measured and simulated data is illustrated in the
14  Supporting Information for the three compounds with the highest concentrations (Figure S2
15  and S3) and indicators for the goodness of fit are summarized in Table 5. In the column
16  study, the measured and simulated concentrations agree reasonably well (Figure S2 and Table
17  5) with correlation coefficients $R^2 \geq 0.69$ using the biodegradation rate constants from a
18  previous study. The average of the absolute difference between measured and simulated $\delta^{13}C$
19  is ≤0.6‰ except for toluene. Measured and simulated concentrations agree less well for the
20  field experience especially for compounds present at low concentrations, which is not
21  surprising given the simplifying assumptions of the model. However, the isotope data are
22  reproduced well with an average absolute deviation of ≤0.9‰ for the three compounds
23  present at the highest concentration (Figure S3, Table 5). Concentration data are more
24  strongly affected by small scale concentration variations and uncertainty introduced by
25  sampling (e.g. small variations in sampling volume) than isotope data. Furthermore,

measured and simulated concentrations might deviate because the source was represented as separate compounds rather than as a multicomponent NAPL.

The isotope evolution varied in space and time and three phases can be observed both in the measured and modelled data. Shortly after source emplacement, depletion in $^{13}$C was observed with increasing distance from the source. The shifts in $\delta^{13}$C were larger for small molecules such as *n*-hexane (-3.8‰ and –5‰ for the column and the field experiment, respectively) compared to larger molecules such as *n*-octane (-2.1‰ and -1.9‰ for the column and the field experiment, respectively). With time, the shifts decreased and $\delta^{13}$C tended to level out to the value in the source. Later, a small enrichment in $^{13}$C with increasing distance from the source was observed for some compounds. Finally, in both experiments, a strong enrichment in $^{13}$C was observed in the vicinity of the source for compounds that were rapidly removed from the source (*n*-hexane and benzene in column study; *n*-hexane and MCP in the field study).

Simulated steady state isotope profiles for *n*-hexane volatilizing from an LNAPL pool across a 3m think unsaturated zone are shown in Figure 4 using three different biodegradation rates (0 d$^{-1}$, 0.1 d$^{-1}$ and 1d$^{-1}$). The simulations were carried out with and without isotope fractionation during diffusion to elucidate the role of diffusion. The results indicate an effect of biodegradation on the isotope ratios, but also demonstrate once more the importance of the diffusion isotope effect. In absence of biodegradation ($k = 0$), the simulations show no shift with distance from the source, even when isotope fractionation due to diffusion is taken into account. For slow biodegradation and a small isotope enrichment factor (-1‰), the compound can become slightly depleted in $^{13}$C with distance despite the occurrence of biodegradation.

The influence of combined diffusion and biodegradation on the isotopic evolution of the source is shown in Figure 5 for the same source scenario using toluene as an example. The source fractionation factor $\alpha_{source}$ was quantified for both stable carbon and hydrogen isotopes using equation 5. For carbon an enrichment of the heavy isotope is predicted while for hydrogen the opposite trend is expected (Figure 5).

## 4 Discussion

### 4.1 $\delta^{13}$C evolution during initial transient phase

The measured and modelled data consistently showed a depletion of $^{13}$C with distance which gradually diminished with time. The isotope fractionation is caused by a smaller diffusion rate of molecules containing a $^{13}$C as indicated by equation 3. The diffusion isotope effect is larger for small molecules as the $^{13}$C contributes more to the total weight of the molecule. In this work, also the hypothetical case was modelled in which diffusion does not fractionate, and thus only biodegradation controlled the isotope fractionation. The isotope ratios were always increasingly enriched in the heavy isotopes with increasing distance from the source (results not shown).

### 4.2 $\delta^{13}$C evolution during steady state

For compounds that show little mass loss from the source over the experimental period, stable steady state concentration and isotope ratio profiles are expected to develop. For compounds with a larger mass loss, steady state may only be reached in approximation and for a short period of time as the source concentration and isotope ratio is steadily shifting. In the following, the isotope profiles during steady state are discussed while profiles during source depletion will be evaluated below (4.3).

Insight into expected isotope profiles at steady state can be gained by considering Figure 4. When biodegradation is absent, constant $\delta^{13}$C values at the level of the source are expected independent of whether or not diffusion is associated with isotope fractionation. In contrast, isotope profiles in presence of biodegradation are affected by diffusion. This is illustrated by comparing steady state isotope profiles for a scenario where diffusion fractionates to a (virtual) scenario where molecules diffuse without isotope fractionation. Fractionation by diffusion counteracts fractionation by biodegradation, especially at a high biodegradation rate, because the preferential removal of $^{12}$C-molecules due to biodegradation is partly counterbalanced by the larger mass flux of $^{12}$C-molecules away from the source. At low biodegradation rates and small isotope enrichment factors, even small shifts in the negative direction with increasing distance might occur. Hence for biodegradation under unsaturated conditions, smaller changes in isotope ratios for a given biodegradation rate and isotope enrichment factor are expected than under saturated conditions. Indeed, modelled and measured isotope ratios in the column study varied little with distance from the source during later periods of the experiment (167 and 336d) despite of the occurrence of biodegradation (Figure 3), except for the two points of n-hexane and methyl-cyclohexane furthest away from the source. These points showed larger shifts than predicated possibly due to spatial variations in the degradation rate, which were not taken into account in the model.

Similar simulations also can be performed in a case where VOCs are out gassing from contaminated groundwater instead of a floating pool. However, the mass transfer limitation between groundwater and soil gas needs to be modelled with great care, since the flux can be either controlled at the groundwater side or in the capillary fringe. The low transverse dispersivity in groundwater creates poor vertical mixing within the saturated zone which may control the volatilisation flux to soil air (Atteia and Höhener, 2010; Parker, 2003). Likewise, the flux of VOCs from the groundwater to the water may be controlled by diffusion in the capillary fringe, and the isotope ratio of the compound in the soil air right above the capillary fringe will be governed by the aqueous-phase diffusion coefficient. A positive offset in the isotopic ratio in the air phase with respect to that in groundwater is then expected because the mass flux of light molecules is higher than that of heavy molecules.

**4.3 $\delta^{13}$C evolution during final transient state due to source depletion**

1   In both experiments, small molecules became enriched in $^{13}$C at or close to the source in the

2   final phase of the vaporisation. Figure 2 and 3 shows the predicted isotope evolution during

3   the final transient state when the source was strongly depleted. At $t = 336$h in the column

4   experiment, simulated $\delta^{13}$C values at the source were -28.3‰ and -25.5‰ for *n*-hexane and

5   benzene respectively, which corresponded well to the measured data of -27.8‰ and -25.6‰,

6   respectively. At $t = 114$ days in the field experiment, simulated $\delta^{13}$C values at the source were

7   –23.3‰, -26.8‰, -25.9‰ and -25.5‰ for *n*-hexane, MCP, MCH and toluene, respectively

8   which corresponded well to the measured data of –21.2‰, -27.1‰, -26.5‰ and –24.2‰,

9   respectively ($x = 0,75$m for hexane and MCP). As indicated by equation 5, the enrichment of

10  heavy carbon isotopes in the source is caused by the mass-dependence of the diffusion

11  coefficient and the isotope effect during biodegradation, while the isotope effect associated

12  with vaporisation can be neglected. Biodegradation influences the source evolution indirectly

13  by creating steeper concentration gradients for light subspecies and hence accelerating the

14  diffusive flux for light subspecies compared to heavy subspecies. This leads to a stronger

15  enrichment of $^{13}$C over time in the source than in absence of biodegradation (Figure 5). For

16  hydrogen isotopes, the normal biodegradation isotope effect partly compensates the inverse

17  isotope effect associated with vaporization that controls the overall hydrogen isotope

18  evolution at the source (Figure 5).

19  **4.4 Assessing biodegradation using isotope data**

20  As discussed in 4.2, diffusion partly counteracts isotope fractionation associated with

21  biodegradation. Because diffusion and biodegradation affects the $\delta^{13}$C in the opposite

22  direction, omission of isotope fractionation by diffusion leads to an underestimation of the

23  amount of biodegradation when interpreting field data based on stable carbon isotope shifts.

24  Furthermore, in a diffusion controlled system, the Rayleigh equation cannot be applied in its

25  original form because the concentration also decreases in absence of degradation. However, it might be possible to relate the decrease in mass flux with distance to the isotope ratio analogously to the Rayleigh equation. Model results show that the change in $\delta^{13}C$ depends on the relative decrease in mass flux on a logarithmic scale (Figure 6). However, the slopes (Table 4), here denoted as effective enrichment factors, are significantly smaller than isotopic enrichment factor for biodegradation of *n*-hexane (-2.3‰) used in the simulations. The smaller isotope enrichment factor can be explained by the fact that the faster diffusion of $^{12}C$-molecules partially compensates the preferential removal of $^{12}C$-molecules by biodegradation.

## 5. Summary and conclusions

The analytical model developed in this study was capable of simulating concentration and isotope data from a column and field study at a reasonable accuracy. The study also provided insight into factors controlling the isotope evolution of VOC distributions under unsaturated conditions. Despite the simplifying assumptions of the model, in the column and field experiments, the average of the absolute difference between measured and simulated $\delta^{13}C$ values was <1‰ for the major compounds. At the field site, a better agreement was observed for isotope than for concentration likely because isotope data are less affected by small scale concentration variations and sampling effects.

Compared to the saturated zone, application of CSIA in the unsaturated zone is more demanding since isotope fractionation by diffusion and vaporization must be taken into account jointly with biodegradation in order to adequately interpret the measured isotope ratios. Based on the simulations, several conclusions regarding the $\delta^{13}C$ behaviour in the unsaturated zone can be drawn. Regarding the use of isotopes to assess biodegradation, a lack of biodegradation can easily be identified since uniform $\delta^{13}C$ values with distance are expected in absence of biodegradation even though $^{13}C$-molecules diffuse at a lower velocity. When biodegradation occurs, a shift in $\delta^{13}C$ occurs which is however smaller than in the

1 saturated zone for a given isotope fractionation factor. Isotope fractionation follows a
2 Rayleigh trend when considering the mass flux decrease rather than the concentration
3 decrease with an effective isotope fractionation factor that is smaller than the isotope
4 fractionation factor for biodegradation only and varies depending on the degradation rate.
5 Finally, the shift towards more positive ratios of the whole profile due to the source depletion
6 can be seen as an indicator of the final stage in the $\delta^{13}C$ evolution and that source exhaustion
7 is imminent.

**Acknowledgment**

We acknowledge financial support provided by the Swiss National Science Foundation and the Swiss Federal Office for Education and Research in the frame of COST Action 629.

**Figure captions:**

**Figure 1:** Schematic representation of a spherical NAPL source retained by capillary forces in the unsaturated zone (a) and a plane NAPL source formed by a large NAPL pool floating on the water table (b).

**Figure 2:** Measured and simulated isotope ratios for the column experiment (measured data from (Bouchard et al., 2008a).

**Figure 3:** Measured and simulated isotope ratios for the field experiment (Measured data from (Bouchard et al., 2008b).

**Figure 4 :** Effect of diffusion and biodegradation (filled symbols) and biodegradation only (open symbols) on the steady state $\delta^{13}C$ profiles of *n*-hexane for three biodegradation rates (0, 0.1 and 1 $d^{-1}$) for a floating pool source above the groundwater table (elevation = 0) and an unsaturated zone of 3m thickness. Unless otherwise indicated, an isotope enrichment factor of -2.3‰ was used.

**Figure 5:** Influence of combined diffusion and biodegradation on the isotopic evolution of toluene volatilising from a floating NAPL source at 3 m below soil surface. The calculation was carried out using the Rayleigh equation and equation 5 defining the isotope fractionation factor. The parameters given in Table 1 were used and the following $^hP/^lP$: 1.0002 for $^{13}C$ (Harrington et al., 1999), 1.005 for $^2H$ (Kiss et al., 1972).

**Figure 6:** Steady state $\delta^{13}C$ values as function of the relative mass flux for a spherical source (A) with $k = 0.1$ $d^{-1}$ and a floating NAPL pool (B) with $k = 1$ $d^{-1}$. The slope corresponds to the net isotope enrichment factor that accounts for isotope fractionation by diffusion and biodegradation.

**Figure 1:**

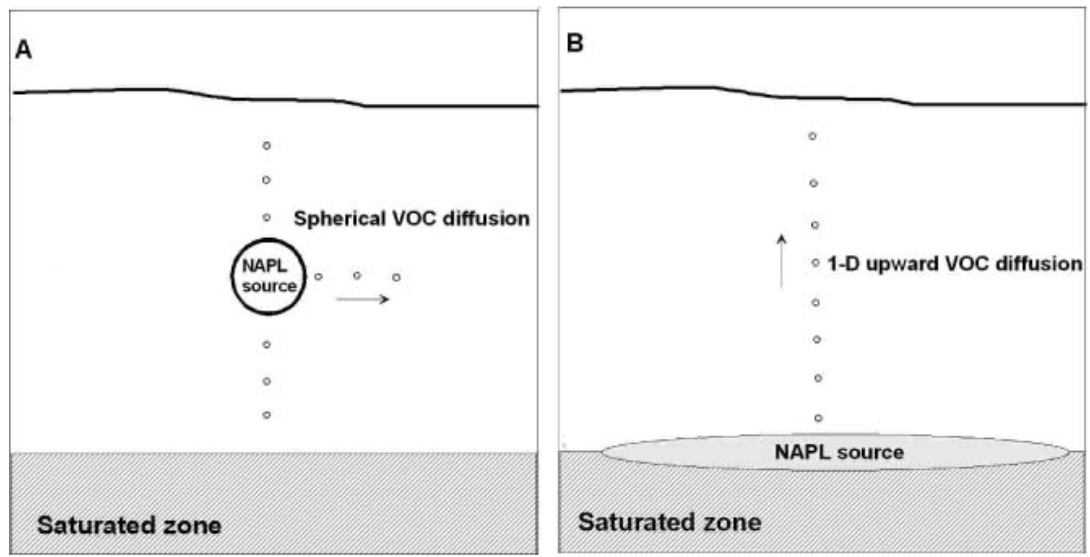

**Figure 2:**

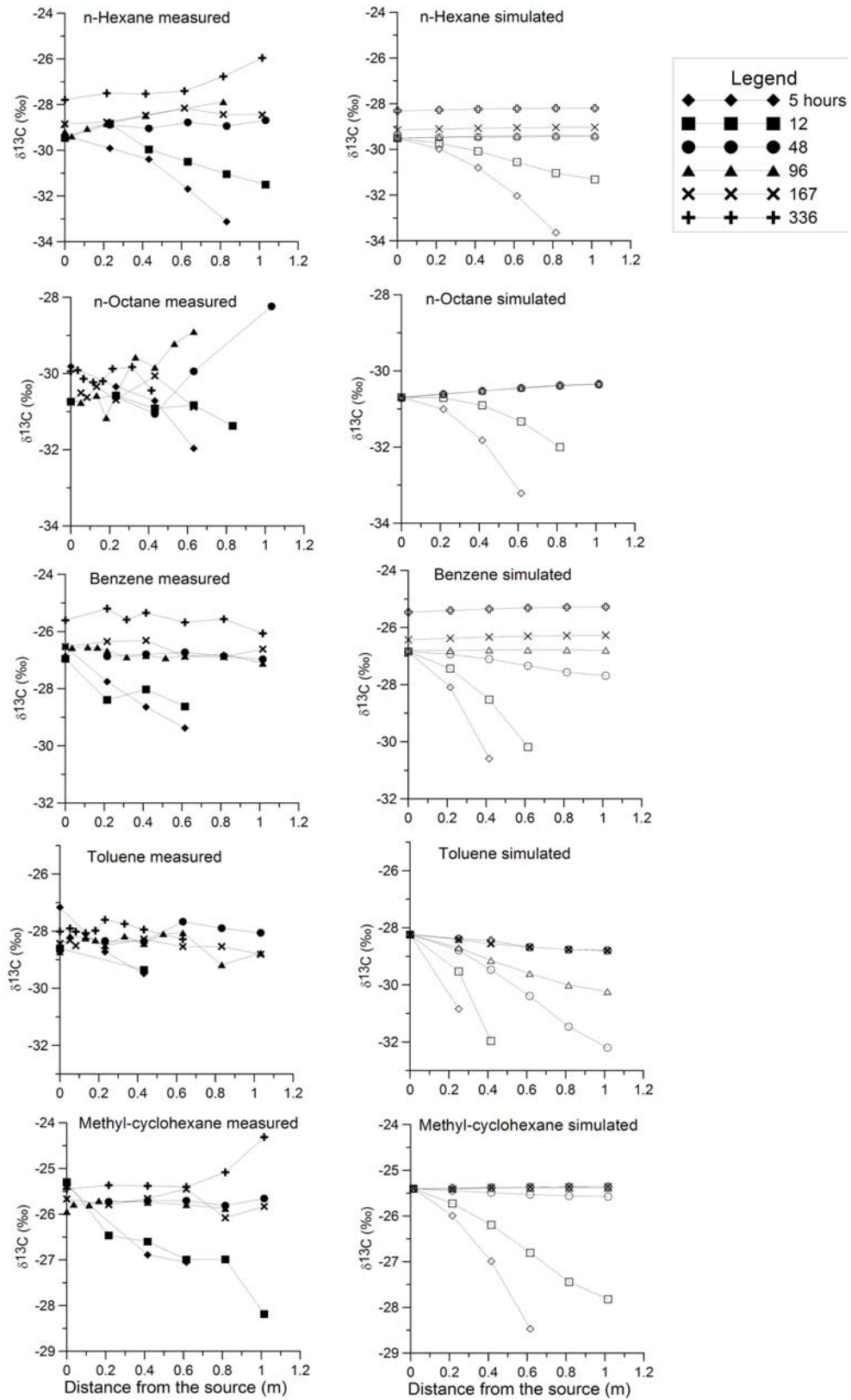

**Figure 3:**

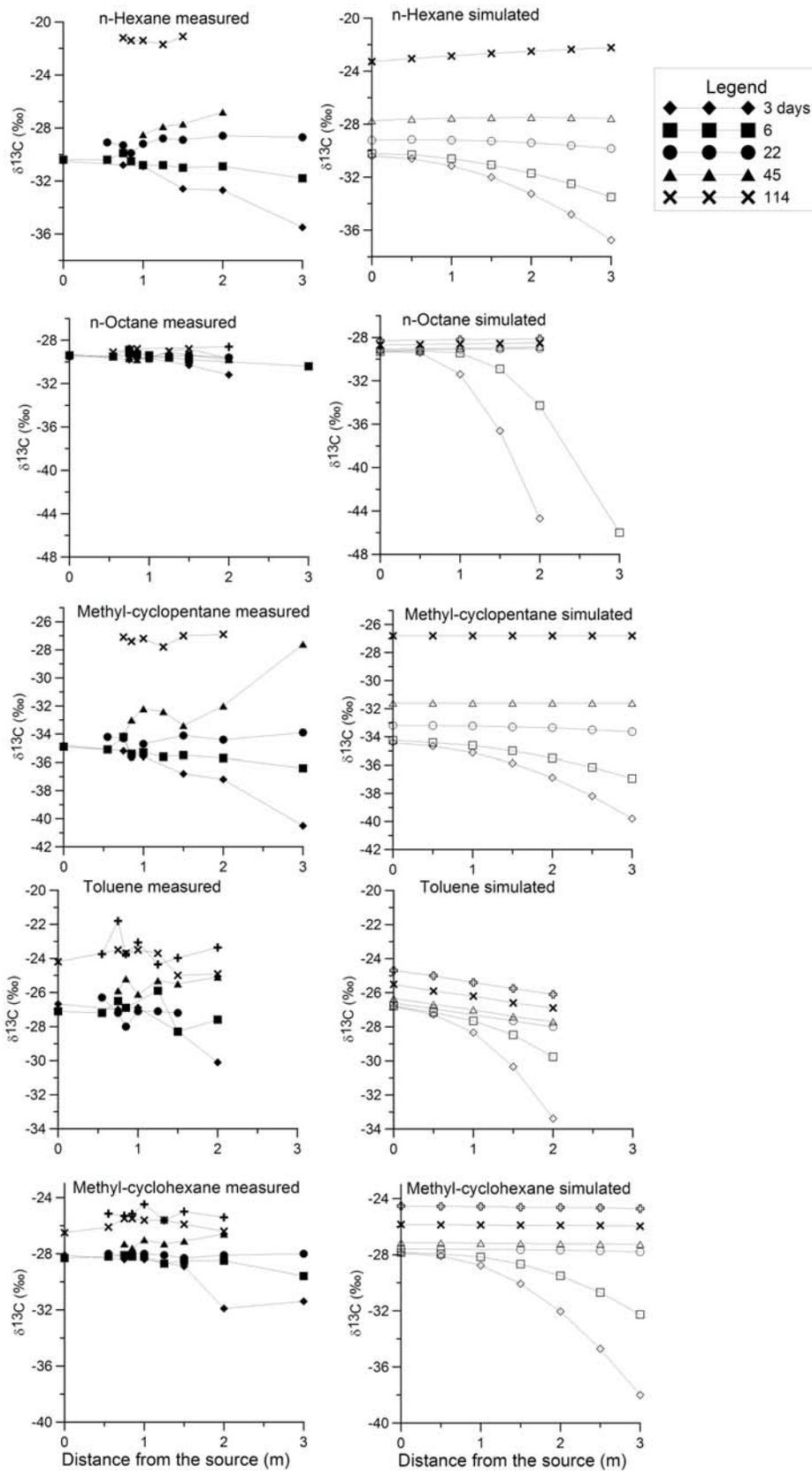

**Figure 4:**

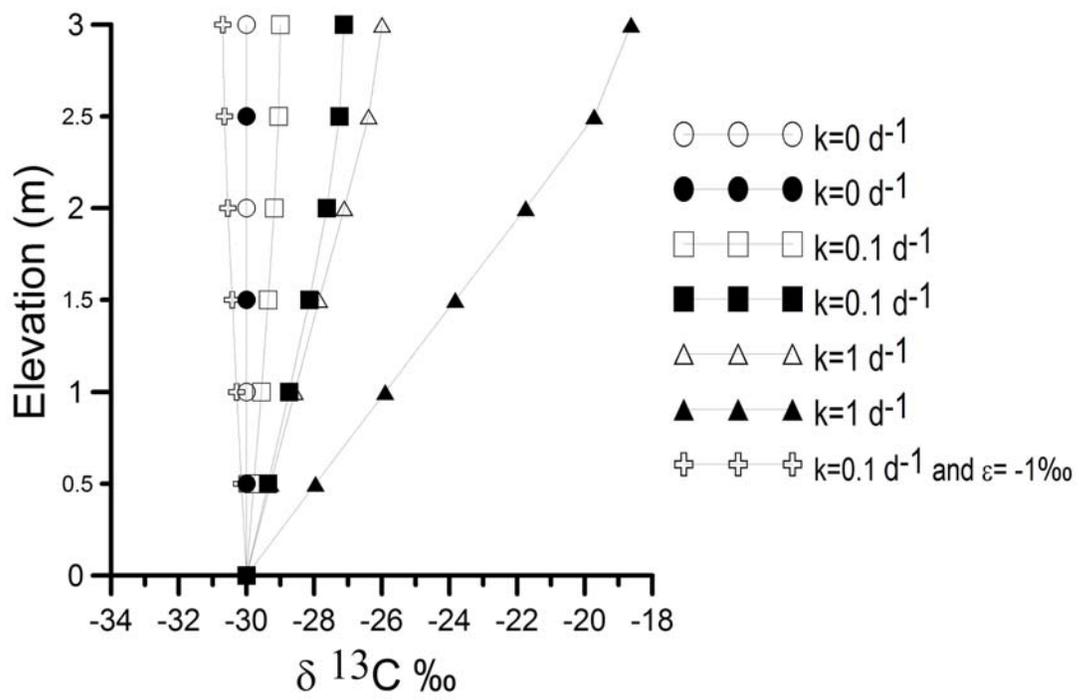

**Figure 5:**

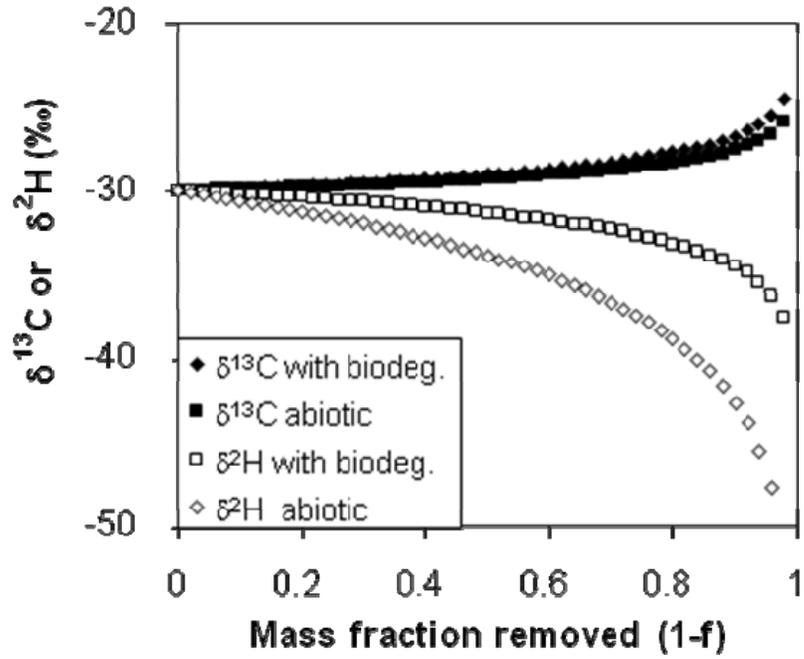

**Figure 6:**

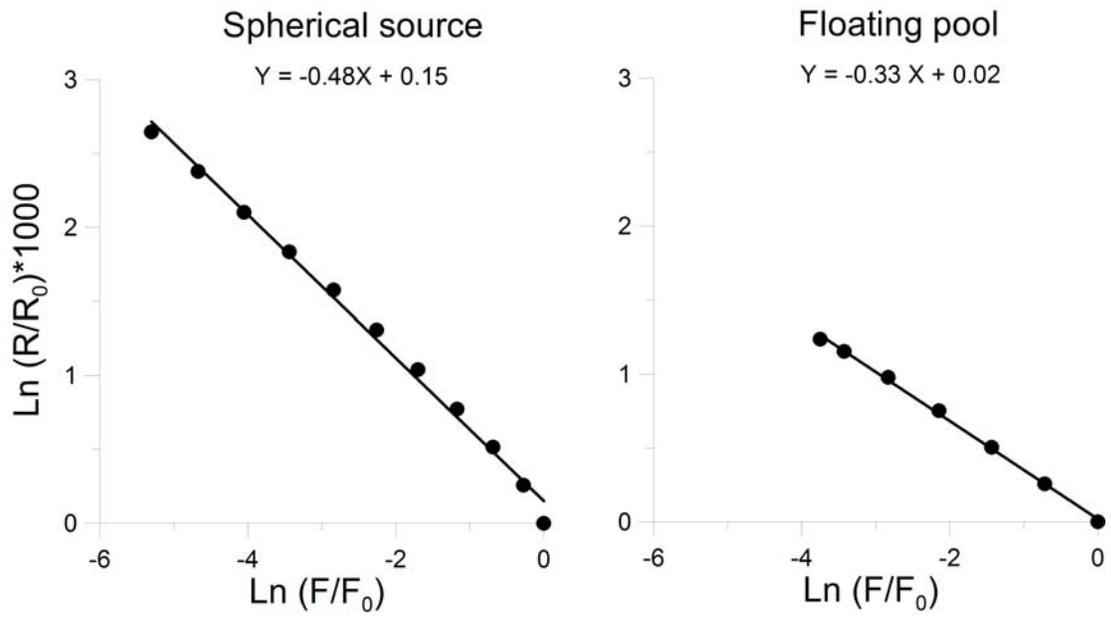

**Table 1:** Values of parameters used for the modelling of selected compounds measured during the column experiment (at 23.5°C): initial concentration ($C_0$), $^{13}C/^{12}C$ ratio ($\delta$), isotope fractionation factor ($\alpha$), sorption-affected diffusion coefficients ($D_s$), apparent biodegradation rate constant ($k$) and source decay rate constant ($\beta$). Using $\omega$ = 96h for *n*-hexane and benzene.

| | $^lC_0$ | $^hC_0$ | $\delta^{13}C$ | $\alpha^1$ | $^lD_s$ | $^hD_s$ | $^lk$ [2] | $^hk$ | $^l\beta$ | $^h\beta$ |
|---|---|---|---|---|---|---|---|---|---|---|
| | (g/m$^3$) | (g/m$^3$) | (‰) | | (m$^2$d$^{-1}$) | (m$^2$d$^{-1}$) | (d$^{-1}$) | (d$^{-1}$) | (d$^{-1}$) | (d$^{-1}$) |
| *n*-Hexane | 72.4368 | 5.0132 | -29.51 | 0.9978 | 0.2791 | 0.2787 | 0.34 | 0.3393 | 0.0812 | 0.08107 |
| *n*-Octane | 7.2373 | 0.6827 | -30.68 | 0.999 | 0.1239 | 0.1238 | 3.25 | 3.2468 | na | na |
| Benzene | 12.6425 | 0.8775 | -26.81 | 0.9979 | 0.1087 | 0.1085 | 0.36 | 0.3592 | 0.0807[3] | 0.08055 |
| Toluene | 5.648 | 0.462 | -28.25 | 0.9992 | 0.0498 | 0.0497 | 0.74 | 0.7394 | na | na |
| | | | -30 [3] | 0.998 [3] | | | | | | |
| MCH | 27.7158 | 2.2742 | -25.38 | 0.9989 | 0.2543 | 0.2539 | 0.11 | 0.1099 | na | na |

[1] from (Bouchard et al., 2008c), MCH from (Bouchard et al., 2005)

[2] value for *n*-pentane from (Hohener et al., 2003)

[3] values used for figure 5

na: not available/ not applicable

**Table 2:** Values of the parameters used for the modelling of selected compounds measured during the field experiment (at 16°C): initial mass of compound ($M_0$), $^{13}C/^{12}C$ ratio ($\delta$), isotope fractionation factor ($\alpha$), sorption-affected diffusion coefficients ($D_s$), apparent biodegradation rate constant ($k$), source decay rate constant ($\beta$) and fraction in the air ($f_a$). Using $\omega = 0$

|  | $^lM_0$[1] | $^hM_0$ | $\delta^{13}C$ | $\alpha$[2] | $^lD_s$ | $^hD_s$ | $^lk$[3] | $^hk$ | $^l\beta$ | $^h\beta$ | $f_a$[4] |
|---|---|---|---|---|---|---|---|---|---|---|---|
|  | (kg) | (kg) | (‰) |  | (m²d⁻¹) | (m²d⁻¹) | (d⁻¹) | (d⁻¹) | (d⁻¹) | (d⁻¹) |  |
| *n*-Hexane | 0.5371 | 0.0371 | -30.6 | 0.9978 | 0.1350 | 0.1348 | 0.054 | 0.0539 | 0.049 | 0.0489 | 0.74 |
| MCP | 0.4183 | 0.0288 | -34.7 | 0.9985 | 0.1321 | 0.1319 | 0.12 | 0.1198 | 0.05 | 0.0499 | 0.94 |
| MCH | 0.7176 | 0.0587 | -28,1 | 0.9989 | 0.0656 | 0.0655 | 0.31 | 0.3096 | 0.018 | 0.0179 | 0.39 |
| *n*-octane | 0.5014 | 0.0474 | -29.3 | 0.9991 | 0.0121 | 0.0120 | 1.08 | 0.0590 | 0.007 | 0.0069 | 0.08 |
| Toluene | 0.2034 | 0.0166 | -27.0 | 0.9992 | 0.0477 | 0.0476 | 0.47 | 0.0469 | 0.01 | 0.0099 | 0.20 |

[1] from (Broholm et al., 2005)

[2] from (Bouchard et al., 2008c), MCP and MCH from (Bouchard et al., 2005)

[3] *n*-hexane, *n*-octane and toluene from (Gaganis et al., 2004). MCP and MCH from (Hohener et al., 2006)

[4] from (Christophersen et al., 2005)

**Table 3:** Values of the soil parameter used for the three different simulations.

|  | $f_{oc}$ (%) | $\theta_t$ | $\theta_w$ | $\tau$ | Temperature (ºC) |
|---|---|---|---|---|---|
| **Column experiment** | 0.2[a] | 0.41[a] | 0.03 | 0.613[b] | 23.5 |
| **Field experiment** | 0.06[c] | 0.32[c] | 0.07[c] | 0.3[c] | 16[c] |
| **Floating LNAPL scenario** | 0.1 | 0.31[d] | 0.06[d] | 0.325[e] | 15 |

$f_{oc}$: fraction of organic carbon

[a] from (Pasteris et al., 2002)

[b] from (Millington and Quirk, 1961)

[c] from (Christophersen et al., 2005)

[d] from (Werner et al., 2005)

[e] from (Moldrup et al., 2000)

**Table 4:** Effective enrichment factors obtained by the relationship $\ln(R/R_0)$ vs $\ln(F/F_0)$ for the spherical source and the floating pool scenarios with three different biodegradation rates.

|  | **Effective enrichment factors (‰)** | | |
| --- | --- | --- | --- |
|  | 0 d$^{-1}$ | 0.1 d$^{-1}$ | 1 d$^{-1}$ |
| Spherical source | 0 | -0.48 | -0.43 |
| Floating pool | 0 | -0.33 | -0.37 |

**Table 5:** Comparison between measured and simulated data. Maximum measured concentration and correlation coefficient for relationship between measured and simulated concentration. Maximum change in $\delta^{13}C$ ($\Delta\delta^{13}C$) and average of the absolute difference between measured and simulated $\delta^{13}C$ (Dev. $\delta^{13}C$)

| Compound | C Nr | Column $C_{max}$ mg/L | Column $R^2$ | Field $C_{max}$ mg/L | Field $R^2$ | Column $\Delta\delta^{13}C$ | Column Dev. $\delta^{13}C$ | Field $\Delta\delta^{13}C$ | Field Dev. $\delta^{13}C$ |
| --- | --- | --- | --- | --- | --- | --- | --- | --- | --- |
| Hexane | 6 | 69.3 | 0.88 | 28.5 | 0.81 | 7.2 | 0.6 | 14.4 | 0.6 |
| MCP | 6 | - | - | 20.5 | 0.60 | - | - | 13.6 | 0.9 |
| Benzene | 6 | 15.7 | 0.82 | - | - | 4.2 | 0.6 |  | - |
| MCH | 7 | 41.2 | 0.80 | 11.0 | 0.66 | 3.9 | 0.4 | 7.4 | 0.6 |
| Toluene | 7 | 8.3 | 0.69 | 3.9 | 0.54 | 2.3 | 1.3 | 6.6 | 1.3 |
| Octane | 8 | 15.1 | 0.75 | 2.0 | 0.36 | 3.7 | 0.6 | 2.4 | 1.5 |